\documentstyle[12pt]{article} 
\newlength{\dinwidth} 
\newlength{\dinmargin} 
\newcommand{\ba}{\begin{array}} 
\newcommand{\ea}{\end{array}} 
\newcommand{\be}{\begin{equation}} 
\newcommand{\ee}{\end{equation}} 
\newcommand{\bee}{\begin{eqnarray}} 
\newcommand{\eee}{\end{eqnarray}} 
 
\def\parallel{| \hskip-0.03cm |} 
 
\newcommand{\tr}{\mbox{Tr}}

\newcommand{\toright}[1]{\smash{
   \mathop{\hbox to 1.3cm {\rightarrowfill}}
    \limits^{#1}}}

\def\Mem{$\,\, \rm M2 \,\,$}
\def\Membar {$\,\,\overline{\rm M2}\,\,$}

\def\tr {\mbox{Tr}}

\def\cp {{\cal P}}
\def\cq {{\cal Q}}

\begin{document} 
\thispagestyle{empty} 
\addtocounter{page}{-1} 
\begin{flushright} 
QMW 99-03\\
SNUST 99-003\\
{\tt hep-th/9903129}\\ 
\end{flushright} 
\vspace*{1.3cm} 
\centerline{\Large \bf Stable Non-BPS Membranes 
on M(atrix) Orientifold~\footnote{ 
Work supported in part by KOSEF Interdisciplinary Research Grant 
98-07-02-07-01-5, KRF International Collaboration Grant, Ministry of 
Education Grant 98-015-D00054, and The Korea Foundation for Advanced 
Studies Faculty Fellowship. }} 
\vspace*{1.2cm} 
\centerline{\large \bf Nakwoo Kim${}^a$, Soo-Jong Rey${}^{b,c}$, 
Jung-Tay Yee${}^b$ } 
\vspace*{0.8cm} 
\centerline{\it ${}^a$ Physics Department, 
              Queen Mary and Westfield College, 
              London E1 4NS UK}
\vskip0.4cm
\centerline{\it ${}^b$ Physics Department, Seoul National University, 
Seoul 151-742 Korea} 
\vskip0.4cm
\centerline{\it ${}^c$ Asia-Pacific Center for Theoretical Physics, 
Seoul 130-012 Korea}
\vskip0.8cm 
\centerline{\tt N.Kim@qmw.ac.uk, \quad sjrey@gravity.snu.ac.kr, 
\quad jungtay@fire.snu.ac.kr } 
\vspace*{1.8cm} 
\centerline{\Large\bf abstract} 
\vspace*{0.5cm} 
Examples of stable, non-BPS M-theory membrane configuration are constructed via 
M(atrix) theory. The stable membranes are localized on O4 or O8 orientifolds, 
which project Chan-Paton gauge bundle of M(atrix) zero-brane partons to 
USp-type. The examples are shown to be consistent with predictions based on 
K-theory analysis. 
\vskip2.1cm 
\leftline{Keywords: M(atrix) theory, non-BPS state, orientifold, K-theory}
\vskip0.3cm
\baselineskip=20pt 
\newpage 
 
\section{Introduction} 
\setcounter{equation}{0} 
At present, M(atrix) theory \cite{bfss} is the only known nonperturbative, 
partonic definition of M-theory, which unifies all perturbative 
superstring theories. For example, it has been found that the M(atrix) theory 
captures successfully the dynamics of supergraviton, compactification on 
lower-dimensional space with or without orbifolds/orientifolds, and 
identification of twisted states thereof. In particular, the theory were 
able to encompass all known BPS states in string and M-theories.  

In this paper, we point out that M(atrix) theory is also able to encompass 
{\sl stable but non-BPS states} as well. We will be illustrating this in the 
simplest context, namely, membrane (M2) in the presence of ${\bf Z}_2$ 
orientifold 
planes, especially, for configurations in which they are parallel each other. 
In this case, two types of orientifold O${}^\pm$ are possible, projecting
onto SO or USp Chan-Paton gauge bundles {\footnote{In this paper, we will be
denoting orientifold $p$-planes with SO (USp) Chan-Paton gauge bundles as 
O$p^-$ (O$p^+$) planes.} for the zero-brane partons.

In M(atrix) theory, the \Mem on an ${\bf Z}_2$ orientifold is described in 
covering space as a pair of \Mem and \Membar configuration, thus breaking
all the supersymmetries completely (unless the membrane is on top of the 
orientifold).  
This is signalled, among others, by the presence of a tachyonic mode in the 
fluctuation spectrum around the \Mem and \Membar configuration. 
Among the fluctuations, only those compatible with the orientifold will
survive. We will thus investigate under what orientifold choices the tachyonic 
mode might be removed out of the fluctuation spectrum. By exploiting projection 
conditions of the Chan-Paton indices, we will be showing that the \Mem on an 
O$4^+$ or O$8^+$ orientifold is a stable, non-BPS configuration.

Recently, there has been progress in classifying stable non-BPS states 
in string theory \cite{sen}-\cite{bergmangimonhorava}. 
Generic non-BPS states are afflicted with tachyon modes,
which leads to an instability for the states to decay or annihilate. As shown
affirmatively by Sen \cite{sen}, the tachyon condensation 
(with or without orientifolds)
leads to plethora of BPS and (stable) non-BPS D-branes, thus shed new 
light on the identification of D-brane charges. Moreover, Witten 
\cite{witten} has shown that possible stable D-brane charge 
configuration formed out of tachyon condensation is most suitably classified 
in terms of K-theory. 
For example, Type I theory is equipped with O$9^-$
and it allows non-BPS D(-1), D0, D7 and D8-branes which can carry ${\bf Z}_2$
charges, in addition to the well-known supersymmetric D1, D5 and D9 branes.
It has been also found that
non-BPS D3 and D4 branes exist as localized states in O$5^-$ plane 
\cite{gukov,hori}. D3-branes
correspond to (unwrapped) D2-branes in Type IIA string theory and, due to 
Bott periodicity, O$9^+$ will also allow non-BPS D3-branes. 
From these results, one would then anticipate existence of a stable, 
non-BPS \Mem configuration located on O$4^+$ or O$8^+$ orientifold with 
USp Chan-Paton gauge bundle. The present work may then be regarded as a 
natural extension of these results within string theory to M-theory defined 
via M(atrix) theory.

This paper is organized as follows. In Section 2, we will briefly recapitulate 
the dynamics of a pair of \Mem-\Membar in M(atrix) theory. We will extend the 
method developed by Aharony and Berkooz \cite{aharonyberkooz}, from which 
the presence of tachyon mode and orientifold projection thereof 
can be formulated in the most transparent way. In Section 3, we will
recapitulate the
definition of the orientifold in M(atrix) theory and the rule of orientifold 
projection to fluctuation spectrum. 
In Section 4, we will study \Mem on O$8^\pm$ orientifold with particular 
attention to the fate of tachyon mode and, using the orientifold projection
rules presented in Section 3, show that O$8^+$ orientifold projects out the 
tachyon mode and hence lead to a stable, non-BPS state.  
In Section 5, we will study \Mem on O$4^\pm$ orientifold and draw the same
conclusion that O$4^+$ orientifold projects out the tachyon mode. 
In Section 6, we will conclude with a brief discussion on comparison with 
K-theoretic analysis, where the \Mem on O$4^+$ or O$8^+$ orientifolds
is known to be a stable, non-BPS state with ${\bf Z}_2$ charges.

\section{\Mem-\Membar Dynamics}
We will begin with, in M(atrix) theory, dynamics of \Mem-\Membar  configuration
drawing particular attention to the tachyonic instability. In fact, the 
dynamics has been studied by Aharony and Berkooz \cite{aharonyberkooz}. 
In this section,we will repeat essential part of their analysis relevant 
for foregoing 
discussions, but in a more transparent notation. The M(atrix) theory 
is a nonperturbative definition of the M-theory, whose action is given by
\be
S = \int dt \, \tr  \left[ \frac{1}{2} ( D_t X^I )^2 +
\frac{1}{4} [X^I,X^J]^2 + \Theta^T D_t \Theta + i \Theta^T \Gamma_I
[X^I,\Theta] \right] \, .
\label{action}
\ee
Here, $X^I, \Theta^a$ are adjoint representations of gauge group $U(N)$, and
$D_t = \partial_t - i [ A_0, \quad ]$ is gauge covariant derivative. 
In M(atrix) theory, a configuration of \Mem-\Membar, separated by
a distance $r$ along 9-th direction, is described by
\be
{X^{\rm I}}_{\rm M \overline{M}} \quad : \quad 
X^1 = \pmatrix{+Q & 0 \cr 0 & +\overline{ Q} } \quad\quad
X^2 = \pmatrix{+P & 0 \cr 0 & -\overline{ P} } \quad\quad
X^9 = \frac{1}{2} \pmatrix{+r & 0 \cr 0 & -r} \, ,
\label{mmbarconfig}
\ee
and all other $X^I$'s and $\Theta^a$'s vanishing. 
Here, $Q, \bar Q, P, \bar P$ are (N $\times$ N) submatrices obeying 
$[Q,P]=[\bar Q, \bar P]=ic {\bf 1}$, 
$c = {\cal O}(1/N)$. It is straightforward to
check that Eq.(\ref{mmbarconfig}) solves the equations of motion derived
from the action, Eq.(\ref{action}).  

As Eq.(\ref{mmbarconfig}) breaks all supersymmetries spontaneously, there
will be corrections to the total energy of \Mem-\Membar configuration. 
At leading order, the correction gives rise to a static, inter-brane 
potential, which can be extracted by applying the Born-Oppenheimer 
approximation -- integrate out small fluctuations around the configuration, 
Eq.(\ref{mmbarconfig}). For the Born-Oppenheimer approximation, relevant parts 
of fluctuations are from off-diagonal submatrices:
\be
X^I = X^{\rm I}_{\rm M\overline{M}} + 
\pmatrix{ 0 & Y^{\rm I} \cr {Y^{\rm I}}^\dag & 0 } ; \qquad
\Theta^a = \pmatrix{0 & \theta^a \cr \theta^{{\rm T}a} & 0}.
\ee
Expanding the potential energy parts in Eq.(\ref{action}) up to 
quadratic order in $Z \equiv (Y^I, \theta^a)$, one obtains various terms 
of the form $\tr ( Z^\dagger_A {\cal O}_1 Z_B {\cal O}_2 )$, where 
${\cal O}_{1,2}$ are generic functions of classical configuration, 
Eq.(\ref{mmbarconfig}), viz. of $Q, \overline{Q}, P, \overline{P}$ and $r$. 
Taking adjoint basis for the fluctuations, $Z_{A,B}$'s, one can
represent these terms 
compactly as
\bee
\tr ( Z_A^\dagger {\cal O}_1 Z_B {\cal O}_2 )
&=& Z^*_{Aij} {\cal O}_{1im} {\cal O}_{2nj} Z_{Bmn} \nonumber \\
&\equiv& Z^*_A ( {\cal O}_1 \otimes {\bf 1} + {\bf 1} \otimes {\cal O}_2^T )
    Z_B.
\nonumber
\eee
Among the quadratic terms, the bosonic fluctuations $Y_3, \cdots, Y_8$ are
in a diagonal form:
\be
- \sum_{\rm I = 3}^8 
Y^\dagger_{\rm I} \left( {\cal Q}^2 + {\cal P}^2 + r^2 \right) Y_{\rm I}
\label{transversefluc}
\ee
where 
\bee
{\cal Q} &=& Q \otimes {\bf 1} - {\bf 1} \otimes \overline{Q}^{\rm T}
\nonumber \\
{\cal P} &=& P \otimes {\bf 1} + {\bf 1} \otimes \overline{P}^{\rm T}
.
\eee

Consistent with $[Q, P] = 
[\overline{Q}, \overline{P}] = i c {\bf 1}$, it is always possible to 
take a realization taking $Q, \overline{Q}$ as symmetric matrices and
$P, \overline{P}$ as antisymmetric ones, so that
\bee
{\cal Q} &=& Q \otimes {\bf 1} - {\bf 1} \otimes \overline{Q}
\nonumber \\
{\cal P} &=& P \otimes {\bf 1} - {\bf 1} \otimes \overline{P}.
\eee

One thus finds that 
Eq.(\ref{transversefluc}) can be interpreted as a quantum mechanical system of
two-particles connected by a spring, where coordinates and conjugate momenta
operators of the two-particles are $Q, \overline{Q}$ and $P, \overline{P}$, 
respectively. Moreover, the fluctuation matrices $Y_{mn}$'s are interpreted 
as `wave functions', where $Q, P$'s and $\overline{Q}, \overline{P}$'s act 
on the Chan-Paton index $m$ and $n$, respectively. Thus, $Y_{mn}$'s are 
two-body wave functions in which the first and the second particles are in 
$m$-th and $n$-th excited states. Moreover, Eq.(6) exhibits clearly that 
degrees of freedom associated with the center of mass of the analog system
consisting of two particles decouple completely, as physically ought 
to be the case.

Using the commutation relation, 
$[{\cal Q}, {\cal P}] = 2 i c {\bf 1}$, one obtains the fluctuation
spectrum of $Y_3, \cdots, Y_8$ as
\be
M^2_\perp (n; r) = r^2 + 2 c (2n + 1), \qquad (n = 0, 1, 2, \cdots).
\ee

For the bosonic fluctuations $Y_{1,2,9}$, the quadratic part is coupled 
one another:
\be
 - (Y^*_1, Y^*_2, Y^*_9)
\pmatrix{ \cp^2 + r^2 & -\cp\cq + 2ic & r \cq  \cr
 - \cq\cp - 2ic   &  \cq^2 + r^2   & r \cp       \cr 
 r \cq &  r \cp  & \cq^2 + \cp^2 \cr }
\pmatrix{Y_1 \cr Y_2 \cr Y_9 } \, .
\label{massmatrix}
\ee
Diagonalizing Eq.(\ref{massmatrix}), one obtains,
in addition to a zero-mode, two fluctuation spectrum
\bee
M^2_{\parallel -}(n;r) &=& r^2 + 2c (2n-1) \, \qquad \qquad
\label{lowesteigenvalue} \\
M^2_{\parallel +}(n;r) &=& r^2 + 2c (2n+3) \,   
\qquad \qquad (n=0,1,2,\cdots).
\nonumber 
\eee
An important point to be noted for later discussion is that, upon diagonalizing
the matrix, Eq.(\ref{massmatrix}), the $n$-th harmonic oscillator eigenstate 
of $Y_9$ is always accompanied by a linear combination of the $(n+1)$-th and 
the $(n-1)$-th eigenstates of $Y_{1,2}$'s. Thus, $Y_1,Y_2$ are even functions 
when $Y_9$ is an odd one, and vice versa.

For fermionic coordinates, the same interpretation as the above goes through
(except that the `wave functions' $\theta^a$ are Grassmannian) and the
fluctuation spectrum is grouped into a pair,
a consequence following from the fact that $\Theta^a$'s are projected by 
$\Gamma_\perp = \gamma_1 \gamma_2$ for each membrane configuration. 
Half of them are
\be
M_{\rm F-}^2(n;r) = r^2 + 2c (2n) \,\,\,\,\,  \qquad \qquad (n=0,1,2, \cdots)
\nonumber\\
\ee 
and the others are
\be
\,\, M_{\rm F+}^2(n;r) = r^2 + 2c (2n+2) \, , \quad\qquad (n=0, 1,2, \cdots)
\nonumber\\
\ee
respectively.

Now, the potential between membrane and anti-membrane arises from summing over 
zero point fluctuation energies, viz. graded sum of the bosonic and fermionic 
mass spectra obtained above: 
\bee
V_{\rm M\overline{M}} (r) &=& {1 \over 2} {\rm STr} M
\nonumber \\
&=& \sum_{m=1}^{\infty}
\left( 
+ \sqrt{r^2 + 2c (2m-3)}
+\sqrt{r^2 + 2c (2m+1)}
+6\sqrt{r^2 + 2c (2m-1)}  
\right.
\nonumber \\
&& \quad \quad
\left.
-4\sqrt{r^2 + 2c (2m-2)}
-4\sqrt{r^2 + 2c (2m)} 
\right) \, .
\eee
This yields the static interaction potential between \Mem and \Membar.
Evidently, when the \Mem and \Membar approaches close, viz. $r < \sqrt{2c}$,
the interaction develops a complex-valued potential energy, a signal of 
tachyonic
instability. Note that the tachyon mode arises from the lowest eigenvalue of 
$Y_{1},Y_2$ fluctuation, see Eq.(\ref{lowesteigenvalue}).

\section{M(atrix) Orientifolds}
In M(atrix) theory, orientifolds are defined as quotient condition on
$X^{\rm I}, \Theta^a$, the moduli space variables parametrizing transverse
spacetime in the discrete light-cone description. To be specific,
we will be considering a spacetime of the form ${\bf R}^{p,1} 
\times {\cal M}^{9-p}/\Gamma$, where $M^{9-p}$ is a smooth manifold and 
$\Gamma = {\bf Z}_2$ is a discrete symmetry group
acting on ${\cal M}^{9-p}$. 
\subsection{\sl M(atrix) ${\bf Z}_2$ Orientifolds}
The M(atrix) orientifold O$p$ is defined 
as a quotient condition on covering space variables of Chan-Paton gauge
bundle U($2N$): 
\bee
X_{\parallel} &=& + M {X_{\parallel}}^{\rm T} M^{-1} \qquad \qquad 
\, ( {\bf R}^{p, \, 1} \qquad {\rm coordinates})
\nonumber \\
X_\perp &=& - M {X_\perp}^{\rm T} M^{-1} \qquad \qquad
( {\cal M}^{9-p} \quad {\rm coordinates})
\nonumber \\
\Theta_a &=& \Gamma_\Omega \, M {\Theta_a}^{\rm T} M^{-1}, 
\eee
Here, $\Gamma_\Omega = \gamma_{9-p} \gamma_{10 - p} \cdots \gamma_9$ is the
product of Dirac gamma matrices of ${\cal M}^{9-p}$. Consistency of the 
quotient condition restricts possible types of the matrix $M$ \cite{kimrey1}.
One then obtains, for a symmetric choice of $M = {\bf 1} \otimes \sigma_1$, 
the projected Chan-Paton bundle is SO$(2N)$, while, for an antisymmetric
choice of $M = {\bf 1} \otimes \sigma_2$, the projected bundle is USp$(2N)$. 
In what follows, we will denote M(atrix) orientifold with SO or 
USp projections as O$p^-$ or O$p^+$, respectively. The variables 
$X_\perp$ and half of $\Theta^a$'s belong to adjoint representation of 
the respective Chan-Paton gauge bundle. The variables $X_{\parallel}$ 
and the other half of $\Theta^a$'s transform as symmetric (antisymmetric)
tensor representation under the SO (USp) gauge bundles, respectively. 
Note that, 
for ${\bf S}^1/{\bf Z}_2$, ${\bf T}^5/{\bf Z}_2$ and ${\bf T}^9/{\bf Z}_2$
compactifications, a consistent choice of the orientifold projection was found 
to be O$8^-$ 
\cite{kachrusilverstein, kimrey1, banksseibergsilverstein}, 
O$4^+$ \cite{kimrey2, smith}, and O$0^-$ \cite{kimrey3, 
ganoretal}, respectively \footnote{Our classification of M(atrix) orientifolds
is slightly different from those used, for example, in \cite{witten, gukov, 
hori}. In our notation, the Chan-Paton gauge bundles refer to M(atrix) 
zero-brane partons on orientifolds. In M(atrix) theory, this notation provides
a more convenient bookeeping.}.  

\subsection{\sl Rules of Orientifold Projection}
Consider, near an orientifold (of transverse distance $r/2$ away), placing
an \Mem parallel to the orientifold plane. In covering space, the 
configuration corresponds to an \Mem-\Membar pair, separated by a distance $r$.

After orientifolding, the off-diagonal submatrices of $X^{\rm I}, \Theta^a$'s
in Eq.(13) will become symmetric or antisymmetric, depending on the choice of 
Chan-Paton gauge bundles. This is because,  
in covering space description, fluctuation modes of $X^{\rm I}, \Theta^a$'s 
that are not consistent with orientifold condition will be projected out
\cite{kimrey1}.
For O$p^-$, it turns out $Y_{\parallel}$'s are symmetric, while $Y_\perp$'s 
are antisymmetric. For O$p^+$, it is the
opposite, viz. $Y_{\parallel}$'s are antisymmetric, whlie $Y_\perp$'s are
symmetric. For $\Theta^a$'s, $\pm$ eigenstates of $\Gamma_\Omega$ are 
symmetric and antisymmetric, respectively. 

In analyzing the fluctuation spectrum, we have interpreted
$Y^{\rm I}$, $\theta^a$'s as quantum-mechanical `wave functions' of an analog
sytem, which consists of two particles connected by a spring. Orientifold 
operation takes a transpose of each 
matrices, see Eq.(13). As the Chan-Paton indices $Y_{mn}$, $\theta_{mn}$ 
are interpreted as that the first harmonic oscillator is in $m$-th excited
state and the second in $n$-th excited state, under orientifold operation,  
the sign of relative coordinates between the two harmonic oscillators gets 
reversed. This implies that, in symmetric submatrices, only even modes 
($n=0,2,4, \cdots$) will survive, while, in antisymmetric submatrices, only 
odd modes ($n=1,3,5,\cdots$) will do so. 

Thus, projection rules of M(atrix) orientifold would be that, 
for O$p^-$ orientifold corresponding to SO Chan-Paton gauge bundle, 
even modes of $Y_{\parallel}$ and odd modes of $Y_\perp$ will only survive.
For $\theta^a$'s, half of even modes and half of odd modes will survive. 
For O$p^+$ orientifold corresponding to USp Chan-Paton gauge bundle, odd
modes of $Y_{\parallel}$ and even modes of $Y_\perp$ will only survive. 
For $\theta^a$'s, again, half of even modes and half of odd modes will 
survive.

Using orientifold projection rules stated as above, we will now analyze
\Mem on O8 and \Mem on O4 configurations in detail.  
\section{\Mem on M(atrix) O8-Orientifold}
Utilizing the fluctuation spectrum of \Mem - \Membar analyzed in section 2, 
we will now examine stability of \Mem located near an O$8$-orientifold
in M(atrix) theory.
\subsection{\sl \Mem Configuratoin Near O8-Orientifold}
We will be taking O8-orientifold spans ${\bf R}^{8,1} = (t, X^1, \cdots, X^8)$.
The \Mem is then located at a distance $r/2$ along $X^9$, the coordinate of 
${\cal M}/{\bf Z}_2$, and extended along $(X^1, X^2)$ directions.
From the supersymmetry transformation rules, one finds easily that the
\Mem-O8 configuration breaks all the supersymmetries of the M(atrix) theory. 

We will now analyze the stability for O$8^\pm$-orientifold projections 
explicitly. This amounts to examining fate of the (complex-valued) tachyon 
mode, the diagonal linear combination of $Y^1,  \, Y^2$ in Eq.(8), under 
the orientifold projection. 

\subsection{\sl \Mem on O$8^-$ Orientifold}
For O$8^-$ orientifold corresponding to SO Chan-Paton gauge bundle, from
the above projection rules, one only keeps even modes of Eq.(7) and Eq.(8)
, and even and odd modes separately for Eqs.(10,11). 

Stability of the \Mem configuration may be examined, for example, from the
static potential when the \Mem is located off the O$8^-$ orientifold at 
a distance $r/2$:
\bee
V_{\rm SO} (r)  &=& 
\sum_{m=1}^{\infty}
\left(  \sqrt{r^2 + 2c (4m-5)} + \sqrt{r^2 + 2c (4m - 1)}
+6\sqrt{r^2 + 2c (4m-3)} 
\right.
\nonumber \\
&&
\quad \quad \left.
 -2\sqrt{r^2 + 2c (4m-4)}
-4\sqrt{r^2 + 2c (4m-2)}
-2\sqrt{r^2 + 2c (4m)} 
\right) \, .
\label{o8-potential}
\eee
Terms in the first line are contribution of even modes of $Y^1, Y^2$, which
will automatically projects into odd modes of $Y^9$ at the same time, and
of even modes of $Y^3, \cdots, Y^8$. Likewise, those in the second line are 
contributions from even and odd modes of $(1 \pm \Gamma_\Omega) \Theta^a$, 
respectively.

At short distance, $r \rightarrow 0$, the potential, Eq.(\ref{o8-potential}),
is complex-valued, as seen from the $m=0$ contribution of the first term
in the summand. This signals a tachyonic instability of the \Mem located
near the O$8^-$-orientifold. To explore $r \rightarrow \infty$ long distance
behavior, it is convenient to reexpress the potential, Eq.(\ref{o8-potential}),
into an integral representation, using the identity:
\be
\sqrt{A} = -\frac{1}{2\sqrt{\pi}} \int_0^\infty \frac{ds}{s^{3/2}} e^{-As}.
\nonumber \\
\ee
One finds  
\be
V_{\rm SO} (r) = 4r - \frac{4}{\sqrt{\pi}} \int \frac{ds}{s^{3/2}} e^{-r^2 s} 
e^{2c s}
\frac{\sinh^4 (c s)}{\sinh (4cs)}.
\ee
Expanding the integrand for small $s$, which is a convergent expansion in 
the limit under consideration, one obtains a long-distance behavior
of the static potential for the \Mem-O$8^-$ orientifold configuration:
\be
\quad
V_{\rm SO} (r) = + 4r -{3 c^3 \over 4 r^5} + 
{\cal O} \left({1 \over r^7} \right), \qquad \qquad
(r \rightarrow \infty).
\ee
The potential is attractive. We thus conclude that, near an O$8^-$ orientifold, 
the \Mem is attracted to the orientifold plane 
and eventually becomes a tachyonic configuration.  
\subsection{\sl \Mem on O$8^+$ Orientifold}
We next turn to O$8^+$ orientifold corresponding to USp Chan-Paton gauge 
bundle. According to the projection rules, one now need to keep 
odd modes of Eq.(7) and Eq.(9), and even and odd modes of 
Eqs.(10,11), respectively. 

Again, we will infer the stability of \Mem configuration conveniently 
by locating it at a distance $r/2$ off the orientifold and measure the static 
potential:
\bee
V_{\rm USp} (r) &=& 
\sum_{m=1}^{\infty}
\left( 
\sqrt{r^2 + 2c (4m-3)}
+\sqrt{r^2 + 2c(4m+1)}
+6\sqrt{r^2 + 2c (4m-1)}  
\right.
\label{s1usp}
\nonumber \\
&&
\left.
\quad  -2\sqrt{r^2 + 2c (4m)}
-4\sqrt{r^2 + 2c (4m-2)}
-2\sqrt{r^2 + 2c (4m-4)} 
\right) \, .
\label{o8+potential}
\eee
Most significantly, one notes that the tachyon mode of Eq.(9) (the $n=0$ mode 
of $M^2_{\parallel -}$) is completely projected out in the background of 
O$8^+$-orientifold. The static potential of \Mem is then well-defined for
$r \rightarrow 0$.  

At long distance, $r \rightarrow \infty$, using the integral representation,
one finds:
\be
V_{\rm USp} (r) = - 4 r - {4 \over \sqrt \pi}
\int_0^\infty {ds \over s^{3/2}} e^{-r^2 s} e^{-2cs}
{\sinh^4 (cs) \over \sinh (4cs)}.
\ee
Expanding the integran for small $s$, one obtains:
\be
V_{\rm USp} (r) = -4r - \frac{3c^3}{4r^5} +{\cal O}(\frac{1}{r^7}) \, ,
\ee
indicating a repulsive long-range force. Thus, we conclude that the
\Mem near an O$8^+$-orientifold exhibits a stable but non-BPS configuration.

\subsection{\sl Compact ${\bf M}^1/ {\bf Z}_2$ Orientifolds and Twisted 
Sector States}
In our discussion so far, we have implicitly assumed that the covering space
${\cal M}$ of the orientifold is noncompact. If ${\cal M}$ is compact, say, 
${\cal M} = {\bf S}_1/{\bf Z}_2$, then in
order to cancel anomalous fluxes carried by the orientifolds, one needs to
introduce a twisted sector in the M(atrix) theory. 
For ${\bf S}_1/{\bf Z}_2, {\bf T}_5/{\bf Z}_2$ and ${\bf T}_9/{\bf Z}_2$ 
cases, complete spectrum of the twisted
sector has been determined previously \cite{kimrey1, kimrey2, kimrey3}. 
One may inquire if the
conclusion in the previous subsections on non-BPS \Mem on M(atrix) 
orientifolds would be modified in case the covering space ${\cal M}$ is 
compact, viz. if the effects of twisted sector spectrum is included.  

For ${\bf S}_1/{\bf Z}_2$, consistency condition of anomalous flux and 
gauge anomaly cancellations fixes uniquely that the two orientifolds (located
at two diagonal points on the ${\bf S}_1$ covering space) are O$8^-$ ones 
and that the twisted sector consists of eight supersymmetry singlet fermions 
on each orientifold. 
The twisted sector fermions, which transform as fundamental representation 
under the SO(2N) Chan-Paton gauge bundle and couples only to $X^9$ minimally.
Inferring $X^9$ from Eq.(2), for {\sl all} twisted sector fermions, fluctuation
energy spectrum will then be $r/2$, independent of the analog 
`harmonic oscillator' excitation level $n$ \footnote{ If each twisted
sector fermion is displaced at locations $m_i$, the fluctuation spectrum is 
changed to $\{ r/2 + m_i \}$ accordingly.}. 

The off-diagonal $Y^{\rm I}, \theta^a$'s submatrices in the untwisted sector 
has $(N \times N)$ degrees of freedom overall. However, for each fluctuation
mode, there are $N$-fold degeneracy, corresponding to the center of mass
motion of the analog `two-particle harmonic oscillator' system. Projecting
these degeneracy out, fluctuation energy spectrum of $Y^{\rm I}, \theta^a$'s 
have effectively $N$-fold degeneracy only. 
The twisted sector fermions belong to fundamental representation of the
Chan-Paton gauge bundle, so their fluctuation energy spectrum is $N$-fold
degenerate, the same as that of untwisted sector. This implies that the
leading-order, linear potential for \Mem on O$8^-$, Eq.(16), is cancelled 
exactly by the contribution of twisted sector fermions, resulting in an
attractive long-range potential. 

Another possibility is that the two orientifolds at ${\bf S}_1/{\bf Z}_2$ 
fixed points are O$8^-$ and O$8^+$ type ones, respectively. In this case,
as the anomalous flux from each orientifold cancels each other, there is no
need to introduce a twisted sector. In this case, \Mem appears a stable state
locally near O$8^+$ orientifold, but as it is transported near to O$8^-$ 
orientifold, \Mem ought to exhibit tachyonic instability. Thus, in this case,
stability of \Mem would depend, in a complicated way, both on
the volume of ${\bf S}_1/{\bf Z}_2$ (viz. distance between O$8^+$ and O$8^-$)
and on location $r/2$ of the \Mem. 

\vskip0.5cm
\subsection{\sl Summary}
\vskip0.3cm 
\Mem is a stable, non-BPS configuration when located on O$8^+$ orientifold,
but exhibit tachyonic instability when located on O$8^-$ orientifold.  
\vskip0.5cm
\section{\Mem on M(atrix) O4-Orientifold}
In this section, we will extend the analysis of section 3 to \Mem located
on O$4$-orientifold in M(atrix) theory. 
\subsection{\sl \Mem versus O4-Orientifold}
We will consider an \Mem extended parallel to an O4-orientifold. The 
O4-orientifold spans ${\bf R}^{4, 1} = (t, X^1, \cdots, X^4)$ and 
an \Mem stretched along $(X^1, X^2)$ directions at a distance $r/2$ along
$X^9$ direction. Thus, in the covering space viewpoint, the configuration 
is given as in Eq.(2), viz. a \Mem-\Membar dipole separated each other by 
a distance $r$. Again, the \Mem-O4 configuration is not a BPS state
as it breaks all of thirty-two supersymmetries of the M(atrix) theory.

We will now analyze stability of the \Mem on O$4^\pm$ orientifold explicitly
following the same analysis as the O$8$ orientifold case. 
\subsection{\sl \Mem on  O$4^-$ Orientifold}
For O$4^-$ orientifold corresponding to SO Chan-Paton gauge bundle of 
zero-brane partons, from the orientifold projection rules, one only keeps 
even modes of $Y^1, \cdots, Y^4$, odd modes of 
$Y^5, \cdots, Y^9$, and even / odd modes for each chiral sector of $\theta^a$'s. 

The potential for \Mem-O$4^-$ is then readily obtained by summing over the 
modes that are left after the orientifold projection: 
\bee
V_{\rm SO} (r) &=& 
\sum_{m=1}^{\infty}
\left( \left\{ \sqrt{r^2 + 2c (4m-5)} + \sqrt{r^2 + 2 c (4m - 1)} \right\}
\right.
\nonumber \\
&& \quad + \left\{2 \sqrt{r^2 + 2c (4m -3)} +4\sqrt{r^2 + 2c (4m-1)} 
\right\}
\nonumber \\
&&
\left. \,\,  -2\sqrt{r^2 + 2c (4m)}
-4\sqrt{r^2 + 2c (4m-2)}
-2\sqrt{r^2 + 2c (4m-4)} 
\right)
.
\eee
Terms in each line come from projection to even modes of $Y^1, Y^2$ for
the first line, which automatically picks up odd modes of $Y^9$, and to even
and odd modes of $Y^3, Y^4$ and of $Y^5, \cdots, Y^8$ respectively for the 
second line. Terms in the third line are from even and odd modes of 
for $(1 \pm \Gamma_\Omega) \Theta^a$, whose decomposition follows from the
fact that $[\Gamma_\perp , \Gamma_\Omega] = 0$ . 

Obviously, for $r \rightarrow 0$, the static potential is complex-valued, 
arising from the presence of a tachyon mode even after the orientifold
projection (the $n=0$ term in Eq.(9)). Thus, \Mem near O$4^-$-orientifold is 
not a stable state. 

Behavior of the static potential for $r \rightarrow \infty$ can be extracted
from an integral representation:
\bee
V_{\rm SO} (r) &=& 
- \frac{1}{2\sqrt{\pi}} \int \frac{ds}{s^{3/2}} e^{-r^2 s} 
\sum_{n=1}^{\infty} e^{-8c s n}
\left( 
e^{10c s} + 5 e^{2c s} + 2 e^{6c s} -
2 -4 e^{4c s} -2  e^{8c s} \right)
\nonumber \\
&=& 
- \frac{2}{\sqrt{\pi}} \int \frac{ds}{s^{3/2}} e^{-r^2 s} 
e^{-c s} \left( e^{4c s} + e^{2 c s} + 2 \right) 
\frac{\sinh^3 (c s)}{\sinh (4cs )}
\eee
Expand the integrand near $s=0$, which is a valid expansion for 
$r \rightarrow \infty$,
\bee
V_{\rm SO}  (r) & = &
- \frac{2}{\sqrt{\pi}} \int \frac{ds}{s^{3/2}} e^{-r^2 s} 
\left( (c s)^2 + \frac{(c s)^3}{2} + {\cal O}((c s)^4)
\right)
\nonumber \\
& = & -\frac{c^2}{r^3} + {\cal O}(\frac{1}{r^5})
\eee
One finds that the potential is attractive. 
Thus, much as in O$8^-$-orientifold, the \Mem 
localized near O$4^-$-orientifold will be attracted toward the 
orientifold and eventually absorbed into it.   

\subsection{\sl \Mem on O$4^+$ Orientifold}
Turning next to O$4^+$ orientifold corresponding to USp Chan-Paton gauge
bundle of zero-brane partons, from the projection rules, one now keeps 
\bee
V_{\rm USp} (r) &=& 
\sum_{m=1}^{\infty}
\left( 
\left\{\sqrt{r^2 + 2 c (4m - 3) } +  \sqrt{r^2 + 2c (4m+1)} \right\}
\right.
\nonumber \\
 && +
\left\{2 \sqrt{r^2 + 2c (4m - 1)} +4 \sqrt{r^2 + 2c (4m-3)} \right\}
\nonumber \\
&& \left. -2\sqrt{r^2 + 2c (4m)}
-4\sqrt{r^2 + 2c (4m-2)}
-2\sqrt{r^2 + 2c (4m-4)} 
\right)
\label{04usp}
\eee
Again, 
terms in the first line come from projection to odd modes of $Y^1, Y^2$,
which automatically projects to even modes of $Y^9$, and those in the second 
line are from projection to odd modes of $Y^3, Y^4$ and even modes of $Y^5, 
\cdots, Y^8$, respectively. 

From Eq.(\ref{04usp}), one clearly sees that the static potential for  
$r \rightarrow 0$ is well-defined, as the tachyon mode, ($n=0$ of Eq.(9)), 
is projected out completely, as in the O$8^+$-orientifold.

For $r \rightarrow \infty$, the static potential can be extracted from the 
integral representation:
\bee
V_{\rm USp} (r) &=& 
- \frac{1}{2\sqrt{\pi}} \int \frac{ds}{s^{3/2}} e^{-r^2 s} 
\sum_{n=1}^{\infty} e^{-8c s n}
\left( 
e^{-2 c s} + 5 e^{6 c s} + 2 e^{2 c s} -
2 -4 e^{4c s} -2  e^{8c s} \right)
\nonumber \\
&=& 
\frac{2}{\sqrt{\pi}} \int \frac{ds}{s^{3/2}} e^{-r^2 s} 
e^{-3 c s} \left( 2 e^{4c s} + e^{2 c s} + 1 \right) 
\frac{\sinh^3 (c s)}{\sinh (4cs )}
\eee
For $r \rightarrow \infty$, one finds 
\bee
V_{\rm USp} (r) & = &
\frac{2}{\sqrt{\pi}} \int \frac{ds}{s^{3/2}} e^{-r^2 s} 
\left( (c s)^2 - \frac{3 (c s)^3}{4} + {\cal O}((c s)^4)
+ \cdots \right)
\nonumber \\
& = & + \frac{c^2}{r^3} + {\cal O}(\frac{1}{r^5}) , 
\eee
exhibiting a repulsive long-range force between \Mem and O$8^+$-orientifold.

\subsection{\sl${\bf T}^5 / {\bf Z}_2$ Orientifolds and Twisted Sector 
States}
If the covering space ${\cal M}_5$ of the orientifold is compact, from
the analysis of section 4.4, one would expect that the stability as well
as static potential of the \Mem configuration becomes modified. The 
modification arises due to interaction between the \Mem and twisted
sector of the orientifold. 

In so far as the twisted sector is concerned, there is one important
difference of the O4-orientifold from the O8-orientifold.
It has actually more to do with peculiarity of the twisted sector in 
O8 orientifold. For example, the twisted sector of the O$8^-$ orientifold 
consists only of fermions that are gauge singlets under the Chan-Paton 
gauge group \cite{kimrey1, kachrusilverstein, banksseibergsilverstein}. 

In contrast, for O$4^+$ orientirfold, the twisted sector is described by 
(dimensional reduction of) chiral multiplets of $d=6, {\cal N}=1$ 
supersymmetric gauge theory.  
Thus, the twisted sector modes are localized at the fixed points and  
do not couple to fields along the orientifold directions, viz. the
excitation energy spectrum is again a function of $r$ alone. 
As such, when summing up the fluctuations, bosonic contribution of the 
twisted sector is cancelled exactly by the fermionic contribution. 
Thus, for \Mem near O4 orientifold, the twisted sector does not lead
to any change to the stability criterion nor to the static potential
energy.  

\subsection{\sl Summary}
\Mem is a stable, non-BPS configuration when located on O$4^+$ orientifold,
but exhibits a tachyonic instability if located on O$4^-$ orientifold.

\section{Comparison with K-Theoretic Classification}
One would like to see whether the stable configuration of non-BPS \Mem on
orientifolds analyzed in the previous section can be understood from an 
independent analysis, for example, K-theory classification. The relevant 
K-theoretic analysis has been already given in \cite{gukov, hori} . According 
to \cite{witten, horava}, the D-brane charges in Type IIA string theory 
takes values in ${\rm K}^{-1}(X)$, the subgroup of ${\rm K} (S^1 \times X)$ 
consisting of trivial elements when restricted to a point on $S^1$. 

For ${\bf Z}_2$ orientifolds, according to the proposal of \cite{hori},
the D-brane charges of D$q$-brane on O$p$-orientifold 
is classified by \footnote{We are grateful to K. Hori for helpful discussions 
on these results.}
\be
{\rm KO} ({\rm D}^{p-q}, \partial {\rm D}^{p-q}) = {\rm KO}^{-(p-q)} ( \cdot)
\ee
for O$p$-orientifold which projects the Chan-Paton gauge bundle of 
D$p$-brane to SO-type, and
\be
{\rm KSp}({\rm D}^{p-q}, \partial {\rm D}^{p-q}) = {\rm KSp}^{-(p-q)} (\cdot)
= {\rm KO}^{-(p-q+4)}(\cdot) 
\ee
for O$p$-orientifold which projects the Chan-Paton gauge bundle of D$p$-brane
to USp-type, respectively. 

Applied to the cases of interest, where
$p-q = 2$ or $6$, one finds  
\bee
p - q=2 \quad : \quad 
{\rm KO}^{-2} (\cdot) &=& {\bf Z}_2 \qquad {\rm for} \quad {\rm SO-projection} 
\nonumber \\
{\rm KSp}^{-2} (\cdot) &=& \,\, 0 \,\, \qquad {\rm for} \quad {\rm USp-projection}
\label{pq2}
\eee
and 
\bee
p - q = 6 \quad : \quad 
{\rm KO}^{-6} (\cdot ) &=& \, \, 0 \,\, \qquad {\rm for} \quad {\rm SO-projection}
\nonumber \\
{\rm KSp}^{-6} 
(\cdot) &=& {\bf Z}_2 \qquad {\rm for} \quad {\rm USp-projection}.
\label{pq6}
\eee

Recall that we have adopted, from M(atrix) theory point of view, a more natural 
notation for classifying orientifold projection to SO or USp Chan-Paton gauge 
bundle was in terms of that of zero-brane partons of the M(atrix) theory 
(See the footnote 3). In the weak coupling limit of the M-theory,
the zero-brane parton is identified with the D0-brane of Type IIA string
theory and, similarly, M-theory orientifolds with Type IIA orientifolds.
One can thus associate, taking into account of the Bott periodicity, the 
M(atrix) O$8^\pm$ orientifolds with Type IIA O8 orientifolds with 
projection to USp(SO) Chan-Paton gauge bundles, and M(atrix) 
O$4^\pm$ orientifolds with 
Type IIA O4 orientifolds with projection to SO (USp) gauge bundles. 

One thus finds that the K-theory classification given in Eqs.(\ref{pq2}, 
\ref{pq6}) matches perfectly with the result we have deduced directly from 
M(atrix) theory in the previous sections, provided the latter is interpreted 
in the weak coupling, Type IIA string theory limit of the M-theory. 

\section{Conclusion}
In this paper, in M(atrix) theory, we have presented a few examples of 
a stable, non-BPS M-brane configuration on an M-orientifold plane. 
We have proposed M(atrix) theory rules for orientifold projection to the 
fluctuation spectrum around a given configuaration and hence for stability
criterion. We have found that the examples found directly using the rules 
are in perfect agreement with the predictions based on K-theoretic 
classification. Our approach is inherently based on M(atrix)
theoretic description of BPS branes and orientifolds and is readily applicable 
to the Type IIB Matrix theory \cite{ikkt} or to the Type I Matrix theory 
\cite{itoyamatokura}. 

Moreover, combining the method developed in this paper with recent work of 
\cite{awata}, where  M(atrix) theory $(p-2)$-brane is constructed as a 
"vortex" configuration of the effective Abelian Higgs model on the 
worldvolume of $p$-brane and 
$\overline{p}$-brane pair, it ought to be possible to explore deeper aspects   
of stable (BPS or not) branes in M-theory. Further results along this 
direction will be reported elsewhere.

\vskip1cm 
We are grateful to B.S. Acharya, M.R. Gaberdiel, K. Hori and A. Sen for 
helpful discussions.

\end{document}